\documentclass[12pt,preprint]{aastex}

\newcommand{\kms}{km\,s$^{-1}$}

\newcommand{\teff}{${\rm T_{eff}}$}
\newcommand{\rjup}{${\rm R_{Jup}}$}
\newcommand{\mearth}{${\rm M_{\earth}}$}
\newcommand{\mjup}{${\rm M_{Jup}}$} 
\newcommand{\rearth}{${\rm R}_{\earth}$}
\newcommand{\gcc}{g\,cm$^{-3}$}
\newcommand{\gcmsq}{g\,cm$^{-2}$}
\newcommand{\cmss}{cm\,s$^{-2}$}
\newcommand{\logl}{log(L/L$_{\odot}$)}
\newcommand{\logg}{log($g$)}
\newcommand{\jh}{($J$-$H$)}
\newcommand{\av}{A$_V$}
\newcommand{\ak}{A$_K$}
\newcommand{\bck}{BC$_K$}

\slugcomment{Accepted to ApJ Letters, 4 September 2007}
\shorttitle{2M1207B} 
\shortauthors{Mamajek \& Meyer}  
\begin{document}

\title{An Improbable Solution to the Underluminosity of 2M1207B:\\ 
A Hot Protoplanet Collision Afterglow}

\author{Eric E. Mamajek} 
\affil{Harvard-Smithsonian Center for Astrophysics, Cambridge, MA, 02138}

\and

\author{Michael R. Meyer} 
\affil{Steward Observatory, The University of Arizona, Tucson, AZ, 85721}

\begin{abstract} We introduce an alternative hypothesis to explain the
very low luminosity of the cool (L-type) companion to the $\sim$25
M$_{Jup}$ $\sim$8~Myr-old brown dwarf 2M1207A.  Recently, Mohanty et
al.  (2007) found that effective temperature estimates for 2M1207B
(1600\,$\pm$\,100 K) are grossly inconsistent with its lying on the
same isochrone as the primary, being a factor of $\sim$10
underluminous at all bands between $I$ (0.8\,$\mu$m) and $L'$
(3.6\,$\mu$m).  Mohanty et al. explain this discrepency by suggesting
that 2M1207B is an 8 M$_{Jup}$ object surrounded by an edge-on disk
comprised of large dust grains producing 2.5$^m$ of achromatic
extinction. We offer an alternative explanation: the apparent flux
reflects the actual source luminosity.  Given the temperature, we
infer a small radius ($\sim$49,000 km), and for a range of plausible
densities, we estimate a mass $<$ M$_{Jup}$.  We suggest that 2M1207B
is a hot protoplanet collision afterglow and show that the radiative
timescale for such an object is $>$ $\sim$1\%\ the age of the
system. If our hypothesis is correct, the surface gravity of 2M1207B
should be an order of magnitude lower than predicted by Mohanty et
al. (2007).
\end{abstract}


\section{Introduction}

While radial velocity surveys and other techniques have yielded over
200 extrasolar planets, there are to date no convincing images of an
extrasolar planet around a star. A few candidate ``planetary mass
objects'' have been identified co-moving with pre-main sequence stars
and young brown dwarfs \citep[e.g. GQ Lup, Oph 162225-240515, etc.;
e.g.][]{Neuhauser05,Jayawardhana06}.  However further observations
have shown these objects to be higher mass \citep[e.g.][]{Luhman07a}.
A possible exception is the companion to 2M1207A \citep[][2MASSW
J1207334-393254 = TWA 26]{Chauvin04}, for which mass estimates have
ranged between $\sim$3-8 \mjup.  Although its mass inferred from
evolutionary models is well below the deuterium--burning limit, it is
unlikely that the object could have formed through classical core
accretion in a circumstellar disk.  Given its estimated mass ratio
($\sim$ 0.2--0.3), and separation ($\sim$ 50 AU) it is thought to have
formed via gravitational fragmentation, similar to a binary star
system \citep{Lodato05}.

The properties of the 2M1207 system are discussed in detail in
\citet{Mohanty07}. The most striking result from that study is that
while both the colors and near-IR spectrum of 2M1207B are consistent
with models of an unreddened $\sim$1600\,K object, the inferred
luminosity for the object is $\sim$2.5 mag (factor of $\sim$10) below
that expected for a $\sim$5-10 Myr object at all wavelengths.  Even
more remarkable, the object is slightly fainter than the observed
sequence of K and L' absolute magnitudes for older field objects with
\teff\, $\simeq$ 1600\,K \citep{Golimowski04}. \citet{Mohanty07} rule
out a handful of simple resolutions to explain the apparent
underluminosity of 2M1207B, and ultimately settle for an unlikely but
testable hypothesis: that the object is obscured by an edge-on disk of
large circumstellar dust grains producing 2.5$^m$ of gray extinction.
In this contribution, we propose an alternative explanation for the
low luminosity of 2M1207B: namely that it has a small radius. In this
scenario 2M1207B is the hot, long-lived afterglow of a recent
collision between two protoplanets \citep[cf.][]{Stevenson87}.  The
observational consequences of protoplanet collisions have been
discussed elsewhere
\citep{Stern94,Zhang03,Anic07}. Here we review the problem of 2M1207B,
introduce a new hypothesis, discuss its merits and deficiencies, and
offer an observational test that can rule it out.

\section{The Luminosity of 2M1207B}

We first review some observational properties of the 2M1207 system.
2M1207A is classified as M8 with emission--line activity
characteristic of T Tauri stars \citep{Gizis02}.  The system appears
to harbor a circumstellar accretion disk as evidenced by broad
H$\alpha$ emission \citep{Mohanty03}, mid-IR excess \citep{Sterzik04},
and outflow activity \citep{Whelan07}.  The motion of 2M1207A is
consistent with membership in the TW Hya Association (TWA), a loose
group of $\sim$20 stars with mean age $\sim$8-Myr-old situated at a
mean distance of $\sim$50\,pc \citep{Webb99,Mamajek05}.  The
optical/near-IR colors of 2M1207A are consistent with no reddening
given its spectral type.  Comparison of its position in the H--R
diagram with theoretical models suggests a mass of $\sim$25\,\mjup\,
and age consistent with its membership in the TWA \citep{Mohanty07}.

2M1207B shares common proper motion and parallax with 2M1207A within
the astrometric uncertainties \citep[milliarcsecond-level measurements
over a few years;][]{Chauvin04,Chauvin05,Song06,Mohanty07}.  The idea
that B could be a foreground or background field L dwarf has been
ruled out.  At a distance\footnote{Our adopted distance to 2M1207 via
the cluster parallax method has increased by 2$\sigma$ compared to
\citet[][53\,$\pm$\,6 pc]{Mamajek05} due to the effects of a improved
proper motion for 2M1207 \citep{Song06}, and a revised estimate of the
TWA group velocity (22.4\,$\pm$\,1 \kms; which follows Mamajek 2005,
but omits the deviant parallax for TWA 9). The updated velocity
increases the distances to the other TWA members in \citet{Mamajek05}
by 7\%.} of 66\,$\pm$\,5 pc, the companion is at a projected
separation of 51\,$\pm$\,4 AU from the primary.  \citet{Mohanty07}
concentrate their analysis of the temperature for 2M1207B on
comparison of data to the Lyon group model atmospheres.  From the
available H-- and K--band spectra, they determine the best fit model
atmosphere has \teff\, =\, 1600 $\pm$ 100 K from the DUSTY grid of
\citet{Allard01}.  A range of surface gravity from \logg\, =
[3.5--4.5] was explored, but the value is poorly constrained.  They
also compare the available photometry of 2M1207B from 0.9--4.0 $\mu$m
with predictions from the DUSTY models, and conclude that they are
consistent with this temperature estimate. While \citet{Leggett01}
demonstrate that DUSTY models are a somewhat poor fit for old late
L-type field dwarfs, \citet{Mohanty07} have shown that they produce
adequate spectral fits for the young objects 2M1207B and AB Pic B.
2M1207B is significantly redder and dustier than typical L dwarfs, as
predicted for low surface gravity objects.

We explore a complementary approach, comparing the spectra published
in \citet{Mohanty07} as well as the available photometry with template
objects drawn from wide field surveys (predominantly older objects).
The low resolution H and K--band spectrum available at
signal--to--noise ratio of 3--10 is morphologically similar to other L
dwarfs suspected of having low gravity: 2MASS J01415823-4633574
\citep[2M J0141;][]{Kirkpatrick06}, 2MASS J22244381-0158521
\citep[][]{Cushing05}, and SDSS J22443167+2043433 \citep[SDSS
J2244;][]{Knapp04}.  All of these objects exhibit weak metal resonance
lines \citep[e.g.][]{Allers07,Gorlova03}, stronger than expected CO
for their spectral type \citep{Cushing05,McLean03}, and unusual
pseudo--continua in the H--band spectra attributed to collision
induced molecular hydrogen absorption \citep[e.g.][]{Borysow97}, all
indications of low surface gravity.  Given the morphological
correspondence between the spectrophotometry of 2M1207B and these
other low gravity L dwarfs, it is reasonable to assume that 2M1207B
has a similar nature.  Further, the spectrum of 2M1207B shows no signs
of CH$_4$ absorption which would indicate \teff\, below 1400 K.

The colors of 2M1207B are very red compared to observed sequences of
field L dwarfs \citep[e.g.][]{Knapp04,Golimowski04}.  The anomalous L
dwarfs listed above also exhibit this behavior which can be attributed
to low gravity \citep[e.g.][]{Burrows06,Allard01}.  Taking the
observed \jh\, colors of 2M1207B and several plausible intrinsic
colors matches, we searched for reddening solutions that would fit the
SED of 2M1207B.  Adopting the colors of 2M J0141 \citep{Kirkpatrick06}
as a low gravity early L template, we derived \av\, $\simeq$ 9 mag,
which matches the $JHK$ photometry well. However, 2M J0141 has \teff\,
=\, 2000\,K \citep{Kirkpatrick06}, and \av\, $\simeq$ 9 mag implies
\ak\, $\simeq$ 1 mag, which would be insufficient to move 2M1207B
above the old dwarf sequence for that \teff.  Alternatively, adopting
the colors of SDSS J2244 \citep{Knapp04} as a late-L low-gravity
template, we arrive at \av\, $\simeq$ 3.7, again reproducing the
colors of 2M1207B within the (rather large) errors from
$\sim$1-4\,$\mu$m. However, the required reddening (\ak\, $\simeq$ 0.4
mag) is insufficient to solve the underluminosity of 2M1207B. If we
assume that 2M1207B has the \jh\, colors of a late--type M dwarf, we
derive \av\, $\simeq$ 11 mag, but cannot reconcile the observed colors
without invoking excess emission in the K and L--bands.  In the limit
of zero extinction, one can find models of extremely low gravity that
fit the SED \citep[e.g.][]{Mohanty07}, so we take that as the simplest
assumption consistent with the observed properties of known L dwarfs
and informed by model atmospheres.

In order to estimate the bolometric luminosity of 2M1207B, we must
also estimate an appropriate bolometric correction to apply to the
observed absolute magnitude.  Given the distance to the source, the
lack of evidence for interstellar reddening, the available photometry,
and the temperature estimate discussed above, we can apply a
bolometric correction to any flux estimate from 0.9--4.0
$\mu$m. \citet{Golimowski04} demonstrates that BC$_K$ varies little as
a function of spectral type for L dwarfs, so we apply the $K$-band BC
to 2M1207B to minimize the uncertainties in the estimate of M$_{bol}$
\citep[\bck\, = 3.25\,$\pm$\,0.14 mag; same as in ][]{Mamajek05}. This
results in a luminosity estimate of \logl\, = --4.54\,$\pm$\,0.10 dex.
For \teff\, = 1600\,$\pm$\,100 K, the DUSTY models predict BC$_K$ =
3.56\, $\pm$\, 0.07 mag, and a luminosity of \logl\, = --4.66\,
$\pm$\, 0.08 dex.  The difference between adopting the empirical or
theoretical BC values is within the errors, so we conservatively adopt
the empirically derived \logl\, as an upper limit.  Given its
luminosity, 2M1207B lies 4--7 times below where expected for an age
range of 5--10 Myr given its inferred temperature range of 1500--1700
K. \citet{Metchev06} derive an effective temperature for the young
($\sim$0.3 Gyr) L/T transition object HD 203030B that is considerably
lower (by $\sim$230 K) than predicted for its spectral type (L7.5).
If 2M1207B were in fact a late-L spectral type with a temperature of
1100 K, we could reconcile its position in the H-R diagram given its
age.  However, this is {\it $\sim$500 K} cooler than the \teff\,
derived from spectral synthesis models according to \citet{Mohanty07}.
\citet{Burrows06} suggest that the temperature of L-T transition
objects should only weakly depend on temperature. Exploring this
solution to the 2M1207B problem requires higher SNR spectra and
further study.

Can the theoretical evolutionary tracks be wrong in luminosity by such
a large factor? As pointed out by \citet{Mohanty07}, another
well-studied young, low-mass binary (AB Pic) exhibits HRD positions
consistent with the age of the group (Tuc-Hor; $\sim$30 Myr).
\citet{Marley07} have argued that the luminosities of young planets
formed through core--accretion are likely to be over-predicted in
models that initially start objects in high entropy states. However,
this argument does not apply if the object formed through
gravitational fragmentation, as has been suggested
\citep{Lodato05}. The \citet{Marley07} models predict that for ages
$>$Myr after accretion has ceased, all planets that formed through
core-accretion with masses of $<$10\,\mjup\, will be colder than
\teff\, $\simeq$ 800\,K and with \logl\, $<$ --5.2 dex. Hence, while
2M1207B is vastly underluminous for its \teff\, compared to the
``hot-start'' evolutionary models of \citet{Chabrier00,Baraffe03}, it
is overluminous and too hot for the fiducial ``cold-start'' core
accretion models for $<$10\,\mjup\, from
\citet{Marley07}. Observationally, there seems to be no trend that
would suggest an error in the evolutionary tracks that could account
for the HRD position of 2M1207B.

As pointed out in \citet{Mohanty07} one cannot reconcile the observed
spectrum and SED of 2M1207B with its apparent low luminosity given the
available models.  Here we take a different approach, adopting the
derived temperature of 1600 K, and postulate that the observed source
flux is a reflection of its actual luminosity.

\section{A Planet-Collision Theory for 2M1207B}

We hypothesize that 2M1207B is the result of a recent collision of two
protoplanets \citep[cf.][]{Stern94}.  Adopting the temperature and
luminosity in \S2, one derives a radius of 48,700\, $\pm$\, 8,800 km
(= 0.68\,$\pm$\,0.12 \rjup, 7.6\,$\pm$\,1.4 \rearth).  For a range of
densities, the inferred mass and gravity are given in Table
\ref{tab:pred}.

We hypothesize that the object is a hot protoplanet of density $\sim$1
g\,cm$^{-3}$. With this radius and density\footnote{The giant planets
in our solar system have bulk densities of $\sim$0.7-1.4 \gcc, and the
known transiting hot Jupiters have densities of $\sim$0.3-1.3 \gcc\,
\citep{Bakos07}. The post-accretion evolutionary track of a 1 \mjup\,
object by \citet{Marley07} has density $\sim$0.5-0.6 \gcc\, in its
first 10 Myr.}, the fiducial planet would have a mass of $M_B$ = 81
M$_{\earth}$ (0.85 M$_{Saturn}$ = 4.7 M$_{Neptune}$).  Throughout this
discussion we will refer to hypothetical protoplanets $B_1$ and $B_2$
which merged to produce body $B$ (2M1207B).  In our solar system, the
largest planetesimals to impact the planets during the late stages of
accretion appear to have had mass ratios of order $\gamma$ =
$M_{B2}$/$M_{B1}$ $\sim$ 0.1, including the bodies responsible for
producing the obliquities of Saturn, Uranus, and Neptune
\citep{Lissauer91}.  Thus we assume that our $\sim$81 \mearth\, object
was the product of a collision between protoplanets with masses
$M_{B1}$ = $(1 + \gamma)^{-1}M_B$ and $M_{B2}$ = $\gamma(1 +
\gamma)^{-1}M_B$, or $M_{B1}$ = 74\,\mearth\, and $M_{B2}$ = 7
\mearth. Following \citet{Wetherill80}, the {\it minimum} impact
velocity for two planets will be their mutual escape velocity, defined
as:
\begin{equation} 
v_{mut}^2 = \frac{2\,G\,(M_{B1} + M_{B2})}{R_{B1} + R_{B2}}
\end{equation}
In the simplified case of identical densities $\rho$ for bodies $B_1$,
$B_2$, and $B$, one can simplify this equation in terms of mass and
radius of the final planet $B$:
\begin{equation}
v_{mut} = 11.2\,{\rm km\,s^{-1}} \left( \frac{M_{B}}{M_{\earth}} \right)^{\frac{1}{2}} \left( \frac{R_B}{R_{\earth}} \right)^{-\frac{1}{2}} \frac{(1 + \gamma)^{\frac{1}{6}}}{(1 + \gamma^{\frac{1}{3}})^\frac{1}{2}} 
\end{equation}
where the last factor is within $<$15\% of unity for all $\gamma$
$\leq$ 10$^{-1}$.  In our fiducial model, the radii of the fiducial
impactors are 3.0 and 6.4 \rearth, respectively. This leads to a
fiducial impact velocity $>$ 30.6\, \kms. Following \citet{Stern94},
one can calculate the radiative timescale of a long-lived afterglow
from the collision of bodies $B_1$ and $B_2$:
\begin{equation}
\tau_{rad}\, \sim\, 0.33\, {\rm Myr} 
\left(\frac{{\rm M_{B2}}}{{\rm M_{\oplus}}}\right)\,
\left(\frac{{\rm R_B}}{{\rm R_{\oplus}}}\right)^{-2}\, 
{\rm T_{1000}^{-4}}\, v^2_{10}
\end{equation}
where $v_{10}$ is the impact velocity of the impactor in units of 10
km\,s$^{-1}$.  $R_B$ is the radius of the planet after collision, and
$R_{\oplus}$ is the radius of Earth.  $T_{1000}$ is the temperature of
the emitting photosphere in units of 1000\,K. We can rewrite the
radiative timescale in terms of the impactor mass ratio
$\gamma$, the properties of the final body $B$, and assuming impact
velocity equals $v_{mut}$:
\begin{equation}
\tau_{rad}\, \sim\, 0.41\, {\rm Myr} 
\left(\frac{{\rm M_{B}}}{{\rm M_{\oplus}}}\right)^2 
\left(\frac{R_B}{R_{\earth}}\right)^{-3}\, 
{\rm T_{1000}^{-4}}\, f
\end{equation}
where $f$ = $\gamma (1 + \gamma)^{-\frac{2}{3}} (1 +
\gamma^\frac{1}{3})^{-1}$.  For $\gamma$ $\ll$ 1, $f$ $\sim$ $\gamma$
(to within $<$40\%\, accuracy for $\gamma$ $<$ 10$^{-1}$).  For our
fiducial model, $\tau_{rad}$ $>$ 59\, kyr or $\sim$1\%\, the age of
the TW Hya association. These radiative timescales are not negligible,
and suggest that a hot afterglow could be visible for an appreciable
fraction of the system lifetime.  In modeling the collision of a
Jupiter and an Earth-like protoplanet, \citet{Zhang03} estimate that
less than 1\%\, of the impact energy is radiated away in the initial
prompt flash.  They further argue that most of the collision-deposited
energy is locked up deep in the post-collision planet, and radiated
over a long timescale as an afterglow that peaks in the IR.  We
propose that 2M1207B is such a long-lived afterglow.

Could a plausible circum(sub)stellar disk form a planetary system with
the required properties?  The formation and collision of two such
large planetesimals at radii $>$ 10 AU in a protoplanetary disk
surrounding a brown dwarf is very unlikely given the mass surface
density and orbital timescales expected \citep{Goldreich04}.  Perhaps
2M1207B formed at smaller radii as the ice--line in the disk of
2M1207A swept through a large range of inner radii from 10 to 0.1 AU
\citep[cf.][]{Kennedy06} as the young brown dwarf evolved \citep[see
][for an alternate scenario]{Boss06}.  If we consider a primordial
disk surrounding 2M1207A that is marginally gravitationally stable
($\frac{M_{disk}}{M_A}$ $\sim$ 0.1) it would have a total gas+dust
mass of 2--3 \mjup.  Adopting the protoplanetary core mass scenario of
\citet[][ see also Lodato et al. 2005]{Ida04}, the time evolution of
the mass of a planet accreting 10$^{18}$ g planetesimals is:
\begin{eqnarray}
\nonumber M_{\rm p}(t)\, \approx\, 8 M_{\oplus} 
\left(\frac{t}{10^6\,\mbox{yr}}\right)^3
\left(\frac{\Sigma_{\rm d}}{10~ \mbox{g cm}^{-2}}\right)^{21/5}\\
\left(\frac{a}{1\,\mbox{AU}}\right)^{-9/5}
\left(\frac{M_{A}}{M_{\odot}}\right)^{1/2} 
\label{eq:coreacc}
\end{eqnarray}
Where $t$ is time, $\Sigma_d$ is disk surface density of solids, $a$
is the orbital distance, and $M_A$ is the mass of 2M1207A.  Assuming a
disk with the above mass, mass surface density profile $\Sigma$
$\propto$ $a^{-1}$, and an outer radius of 15 AU, the total disk mass
surface density at 3 AU is 250 \gcmsq.  If we consider a gas to
dust+plus ice ratio of 25 (100/4), we arrive at a mass surface density
in solids of 10 \gcmsq\, at 3 AU.  Using the above equation for a
brown dwarf of mass 0.025 M${\odot}$, we estimate that a core of 5-10
\mearth\, can form within $\sim$3 Myr at this radius.  In 10 Myr, a
similar mass core could form at a distance of 5 AU from the brown
dwarf.  Assuming the above disk model parameters, roughly half of the
total disk mass (solid and gas) resides inside of 7.5 AU and half
outside.  It is at least plausible that 2--4 cores of 5--10 \mearth\,
could form between 1--10 AU within 3--10 Myr in this system utilizing
the bulk of available solids in the system.  If two of those cores
accreted enough gas to form Neptune-to-Saturn mass protoplanets, we
can envision a scenario where: a) failed cores of 5--10 \mearth\,
could collide with a successfully formed gas/ice giant protoplanet,
creating the observed hot collisional afterglow; and b) another gas/ice
giant, along with the presence of the remnant primordial disk, could
eject 2M1207B to its observed orbital radius of 50 AU.  Motivated by
evidence for gas giants at large separations having created observed
structure in debris disks, \citet{Veras04} have investigated gas giant
migration/ejection scenarios in disks.  \citet{Thommes03} have also
proposed that Neptune and Uranus formed closer to the Sun (between
Jupiter and Saturn) and were ejected to larger orbital radii through
dynamical processes.  We note that the remnant disk surrounding
2M1207A has a mass comparable to that we propose for the ejected
2M1207B \citep{Riaz07} though its outer radius is unconstrained from
current observations. This is, of course, a highly improbable series
of events.

\section{Predictions} 

The hypotheses of whether or not 2M1207B is a hot protoplanet
collision afterglow or is obscured by a dense disk of large dust grains
can be tested.  In the scenario proposed here, 2M1207B is actually a
$\sim$80\,\mearth\, object with radius $\sim$49,000\,km. The surface
gravity of such an object in cgs units would be log($g$) $\sim$ 3
(Table 1).  This is significantly lower than than for 5-10 Myr-old
3--8 \mjup\, objects which have \logg\, $\approx$ 4 \citep{Mohanty07}.
If 2M1207B possesses an edge-on disk exhibiting grey extinction, then
spatially resolved ground-based observations should reveal: a) an
infrared excess at $\lambda$ $>$ 4 $\mu$m from the disk; b) a 10
$\mu$m silicate absorption feature consistent with the disk being
edge-on; c) polarized emission from scattered light at shorter
wavelengths; and/or d) resolved scattered light emission consistent
with an edge-on dust disk system \citep[cf.][]{Luhman07b}.

If 2M1207B is actually a physically smaller (and therefore lower mass)
companion, it should exhibit near-infrared spectra: a) consistent with
the 1600 K temperature advocated by \citet{Mohanty07}; and b) low
surface gravity (log(g) $\sim$ 3) in high S/N spectra.  As mentioned
above, surface gravity affects the spectra of very cool objects in
ways that can be observed through analysis of atomic and molecular
features.  \citet[][ see also Kirkpatrick et al. 2006]{Gorlova03}
suggest that \logg\, can be estimated to within 0.3-0.5 dex from high
S/N near-infrared spectra of M and L dwarfs.  \citet{Allers07}
specifically investigate the gravity dependence of the \ion{Na}{1}
feature at 1.14 \micron\, while Gorlova et al. provide a preliminary
calibration of the surface gravity effects of \ion{K}{1} at 1.25
\micron\, \citep[see also][]{McGovern04}.  These effects should be
clear in modest S/N spectra (20-30) easily distinguishing between the
\logg\, $\sim$ 4 model of \citet{Mohanty07} and the \logg\, $\sim$ 3
model proposed here.  Further, we anticipate that our protoplanetary
collision remnant would be metal-rich compared to the primary.  Models
from \citet{Burrows06} as well as \citet{Fortney06} demonstrate the
significant differences in brown dwarf and gas giant planet
atmospheric models by varying the metallicity.  Such effects would be
easily observable in S/N $\sim$ 20--30 spectra, obtainable in 1-2
nights of observing time on a 6-10 meter telescope equipped with
adaptive optics. Perhaps future surveys will uncover additional hot
protoplanet collision afterglow candidates with even smaller inferred
masses.

\acknowledgments

EEM is supported through a Clay Postdoctoral Fellowship from the
Smithsonian Astrophysical Observatory. MRM is supported by LAPLACE.
We thank Phil Hinz, Scott Kenyon, Matt Kenworthy, Stan Metchev, Subu
Mohanty, and Steve Strom for discussions. We thank Adam Burrows for
bringing the \citet{Stern94} paper to our attention, Kevin Zahnle for
an inspiring colloquium regarding the Earth-Moon impact, and the
anonymous referee for a thoughtful review.

%% PLACE TABLE 1 HERE

\begin{deluxetable}{crcc}
\tablecolumns{4}
\tablewidth{0pc}
\tablecaption{Predicted Quantities for 2M1207B \label{tab:pred}}
\tablehead{
\colhead{$\rho$} & \colhead{Mass}      & \colhead{Mass}    & \colhead{\logg}\\
\colhead{(\gcc)} & \colhead{(\mearth)} & \colhead{(\mjup)} & \colhead{(\cmss)}}
\startdata
0.5 &  40.5 & 0.13 & 2.83\\ 
1.0 &  81.0 & 0.25 & 3.13\\
1.5 & 121.5 & 0.38 & 3.31\\
2.0 & 162.0 & 0.51 & 3.43
\enddata
\end{deluxetable}

\end{document}